\documentclass[twocolumn,prl,preprintnumbers,amsmath,amssymb,superscriptaddress]{revtex4}

\usepackage{graphicx}
\usepackage{dcolumn}
\usepackage{bm}
\usepackage{array}
\usepackage{mathrsfs}
\usepackage{physics}
\usepackage{times}
\usepackage{float}
\usepackage{booktabs}
\usepackage{autobreak}

\usepackage[colorlinks=true,linktocpage=true,linkcolor=blue,citecolor=blue]{hyperref}

\graphicspath{{figures/}}
\newcolumntype{I}{!{\vrule width 1pt}}
\newlength
\savedwidth

\newlength\savewidth

\begin{document}

\title{ Vector meson spectral function in a dynamical AdS/QCD model}

\author{Yan-Qing Zhao}
\email{zhaoyanqing@mails.ccnu.edu.cn}
\affiliation{Institute of Particle Physics and Key Laboratory of Quark and Lepton Physics (MOS),
Central China Normal University, Wuhan 430079, China}

\author{Defu Hou}
\thanks{Corresponding author}
\email{houdf@mail.ccnu.edu.cn}
\affiliation{Institute of Particle Physics and Key Laboratory of Quark and Lepton Physics (MOS),
Central China Normal University, Wuhan 430079, China}

\date{\today}

\begin{abstract}
By using gauge/gravity duality, we calculate the spectral function of the heavy vector mesons with the presence of an intense magnetic field in a hot and dense medium. The results show that, a general conclusion, as the increases of  magnetic field, chemical potential and temperature, the height of the peak of the spectral function decreases and the width increases. A nontrivial result is the change from the peak position of spectral function. We explain this non-trivial  behavior by the interplay of the interaction between the two heavy quarks and  the interaction between the medium with each of the heavy quarks.

\end{abstract}


\maketitle
\section{Introduction}\label{sec:01_intro}
A new matter state, quark-gluon plasma, has been generated at the Relativistic Heavy-Ion Collider (RHIC) in Brookhaven. These substances are interconnected by strong interactions which can be elaborated by quantum chromodynamics (QCD). So this new state of matter is also  extensively called the strongly-correlated quark-gluon plasma (sQGP)~\cite{Gyulassy:2004zy}. The strength of the coupling between them is related to the energy. At high energy, the coupling is very small, and quarks have the property of asymptotic freedom, which can be studied by perturbation theory. At low energy, the coupling is very large, and quarks have the property of color confinement, so the perturbation theory is no longer applicable and a non-perturbation method needs to be employed. Fortunately, gauge/gravity correspondence or the bulk/boundary correspondence ~\cite{Maldacena:1997re,Witten:1998qj,deHaro:2000vlm,Rey:1998ik,Klebanov:1999tb,Karch:2002sh,Erlich:2005qh,Freedman:1998tz,Policastro:2002se,CasalderreySolana:2011us,Gopakumar:1998ki,Emparan:1999pm,Caldarelli:1999xj,Hawking:1998kw}, a new approach, provides a powerful tool to study the non-perturbative regime of the physic system. The most attractive achievement of the gauge/gravity correspondence is that one can also perform an analytical calculation by this technique.

The suppression of $J/\Psi$ and $\Upsilon(1S)$ peaks in the dilepton spectrum plays an important role in many signals of phase transition. Heavy quarkonium may survive, due to Coulomb attraction between quark and antiquark, as bound states when the quark matter encounters phase transitions. The seminal paper of quarkonium suppression is from  Matsui and Satz~\cite{Matsui:1986dk} where the estimate for the dissociation temperature was first made, $T_{dis}\sim1.2T_c$. By analyzing the spectral function in the Nambu-Jona-Lasinio model, Ref.~\cite{Hatsuda:1985eb} points out that quark-antiquark excitations, small mass and narrow width in the $\pi-\sigma$ channels, still exist at $T>T_c$ where $T_c$ is the phase transition temperature. However, Ref.~\cite{Asakawa:2003re} studies $J/\Psi$ and $\eta_c$,  by discussing the correlation function of $J/\Psi$ at finite temperature on $32^3*(32-96)$ anisotropic lattices using the maximum entropy method (MEM) , in the deconfined plasma from lattice QCD, the results show that $J/\Psi$ and $\eta_c$ can survive in the plasma with obvious resonance state up to $T\simeq1.6T_c$ and  dissociate at $1.6T_c\leq T\leq1.9T_c$. Next, Ref.~\cite{Datta:2003ww} considers the $Q\overline{Q}$ with quark masses close to the charm mass on very fine isotropic lattices, the results show that the $J/\Psi$ and $\eta_c$ can survive up to $1.5T_c$ and vanish at $3T_c$.

RHIC experiments show that sQGP not only has the properties of high temperature and density, but also a very intense magnetic field engendered in noncentral relativistic heavy-ion collision and exists long enough to
also affect the produced quark-gluon plasma ~\cite{Skokov:2009qp,Bzdak:2011yy,Zhu:2019igg,Zhu:2021ucv,Voronyuk:2011jd,Roy:2017yvg,Tuchin:2013ie,Seiberg:1994pq,Gusynin:1995nb}. In recent years, a lot of work has revealed the dependence of thermal QGP properties on the magnetic field and chemical potential~\cite{Kharzeev:2010gd,Cao:2020ryx,Preis:2010cq,Albash:2007bk,Kalaydzhyan:2011vx,Cao:2021tcr,Finazzo:2016mhm,Preis:2012fh,Rougemont:2015oea,Bu:2012mq,Hoyos:2011us,Gorsky:2010xu,Cai:2013pda,Mamo:2015dea,Jensen:2010vd,Dudal:2015wfn,Jokela:2012vn,Callebaut:2011ab,Filev:2010pm,Gursoy:2016ofp,Evans:2016jzo,Braga:2018zlu,Colangelo:2013ila,Gursoy:2017wzz,Braga:2019yeh,Zhu:2019ujc,Li:2016gfn,Zhou:2020ssi,Feng:2019boe}. Ref.~\cite{Fujita:2009wc} deliberates the finite-temperature influences on the spectral function in the vector channel and pointed out that the dissociation of the heavy vector meson tower onto the AdS black hole results in the in-medium mass shift and the width broadening. The lattice QCD studies, for the $J/\psi$ spectral function at finite temperature, show that mesons can survive beyond $2T_c$ where $T_c$ devotes to the phase transition temperature~\cite{Umeda:2000ym,Asakawa:2003re,Datta:2003ww}. Ref.\cite{Ghoroku:2005vt} calculates low-lying (axial-)vector meson mass spectra in a five-dimensional deformed $AdS_5$ holographic QCD by a bulk scalar field, which is a kind of soft-wall model. They pointed out that such deformation effects are important to reproduce the experimental data for vector mesons and the axial-vector mesons. The paper~\cite{Avila:2020ved}, by introducing the probe D7-branes to the 10d gravitational backgrounds to add flavor degrees of freedom, indicates that the magnetic field decreases the mass gap of the meson spectrum along with their masses. Ref.~\cite{Braga:2018zlu} displays the heavy meson dissociation in a plasma with magnetic fields and states clearly the dissociation effect increases with the magnetic field for magnetic fields parallel and perpendicular, respectively, to the polarization. The case of finite density is studied in ~\cite{Braga:2017bml}, which gives that increasing temperature and chemical potential decrease the peak of spectral function showing the thermal dissociation process. Ref.~\cite{Hou:2007uk} calculates the dissociation temperature of $J/\Psi$ and $\Upsilon(1S)$ state by solving the Schr\"{o}dinger equation of the potential model. They find the ratios of the dissociation temperatures for quarkonium with the $U$-ansatz of the potential to the deconfinement temperature to agree with the lattice results within a factor of two.

Inspired by ~\cite{Braga:2017bml,Braga:2018zlu}, the main objective of this manuscript is to investigate the spectral function for $J/\psi$ and $\Upsilon(1S)$ state by considering a more general situation, including chemical potential and magnetic field in gravity background, which could better simulate the sQGP environment produced by RHIC experiments. The introduced magnetic field breaks the $SO(3)$ invariance into $SO(2)$ invariance, that is, the magnetic field only changes the geometry perpendicular to it which is different from other models. The manuscript is organized as follows. In section~\ref{sec:02}, we review the magnetized holographic QCD model with running dilaton and chemical potential. In section~\ref{sec:03}, we give the specific derivation process for calculating the spectral function of $J/\psi$ and $\Upsilon(1S)$ states. In section~\ref{sec:04}, we show and discuss the figure results.

\section{Holographic QCD model}\label{sec:02}
We consider a 5d EMD gravity system with two Maxwell fields and a neutral dilatonic scalar field as a thermal background for the corresponding hot, dense and magnetized QCD, where the finite temperature effects are introduced by black hole thermodynamics, the finite density effects are related to the charge of the black hole, and the real physical magnetic fields are related to the five-dimensional magnetic field. The 5-d Lagrangian is given by
\begin{align}\label{eq01}
 \begin{autobreak}
  \mathcal{L} =
  \sqrt{-g}(R-\frac{g_1(\phi_0)}{4}F_{(1)\mu\nu}F^{\mu\nu}-\frac{g_2(\phi_0)}{4}F_{(2)\mu\nu}F^{\mu\nu}
  -\frac{1}{2}\partial_\mu\phi_0\partial^\mu\phi_0-V(\phi_0)).
  \end{autobreak}
\end{align}
where $F_{(i)\mu\nu}$ ($i=1,2$) is the field strength tensor for U(1) gauge field, $g_i(\phi_0)$ ($i=1,2$) is the gauge coupling kinetic function, $\phi_0$ is the dilaton field, $V(\phi_0)$ is the potential of the $\phi_0$ (see\cite{Bohra:2019ebj} for exact expression). By introducing external magnetic field in $x_1$ direction  damages the $SO(3)$ invariance to $SO(2)$ invariance in boundary spatial coordinates, so we consider the following anisotropic ansatz for the background blackening metric $f_{\mu\nu}$ and $(\phi_0,A_\mu)$ fields in Einstein frame~\cite{Bohra:2019ebj}
\begin{align}\label{eq02}
 ds^2 &= \frac{R^2S(z)}{z^2}(-f(z)dt^2+dx_1^2+e^{B^2z^2}(dx_2^2+dx_3^2)+\frac{dz^2}{f(z)}), \notag \\
\phi_0 & =\phi_0(z),\quad A_{(1)\mu}=A_t(z)\delta_\mu^t,\quad F_{(2)}=B dx_2\wedge dx_3,
\end{align}
with
\begin{align}\label{eq03}
 f(z) &  =1+\int_0^zd\xi \xi^3e^{-B^2\xi^2-3A(\xi)}[K+\frac{\widetilde{\mu}^2}{2R_{gg}R^2}e^{R_{gg}\xi^2}],\notag \\
  K & = -\frac{1+\frac{\widetilde{\mu}^2}{2R_{gg}R^2}\int_0^{z_h}d\xi \xi^3e^{-B^2\xi^2-3A(\xi)+R_{gg}\xi^2}}{\int_0^{z_h}d\xi \xi^3e^{-B^2\xi^2-3A(\xi)}},\notag\\
  \widetilde{\mu} &= \frac{\mu}{\int_0^{z_h}d\xi\frac{\xi e^{-B^2\xi^2}}{g_1(\xi)\sqrt{S(\xi)}}},
\end{align}
where $R$ is the radius of the asymptotic AdS spaces, $S(z)$ is the scale factor, $f(z)$ represents the blackening function and $\mu$ is the chemical potential. We stipulate that the asymptotic boundary is at $z=0$ and $z=z_h$ labels the location of horizon where $f(z_h)=0$.  The concrete form of gauge coupling function $g_1$ can be determined by fitting the vector meson mass spectrum. The linear Regge trajectories for $B=0$ can be restored when
\begin{equation}\label{eq04}
  g_1(z)=\frac{e^{-R_{gg}z^2-B^2z^2}}{\sqrt{S(z)}}.
\end{equation}
Here, this $B$, in units $GeV$, is the 5d magnetic field. The 4d physical magnetic field is $e\mathcal{B}\sim \frac{const}{R}\times B$ where $R=1GeV^{-1}$ is the AdS length and $const=1.6$ (See details in Ref.\cite{Dudal:2015wfn}). For convenience, we will use the 5-Dimensional magnetic field $B$ to participate in the calculation in the subsequent parts. Using $S(z)=e^{2A(z)}$, one can obtain $R_{gg}=1.16GeV^2$ for heavy meson state $J/\psi$. In the following calculation, we take $A(z)=-az^2$ where $a=0.15GeV^2$ matching with the lattice QCD deconfinement temperature at
$B=0GeV$ ~\cite{Dudal:2017max}.

The Hawking temperature with magnetic field and chemical potential can be computed by surface gravity
\begin{equation}\label{eq05}
  T(z_h,\mu,B)=\frac{-z_h^3e^{-3A(z_h)-B^2z_h^2}}{4\pi}(K+\frac{\widetilde{\mu}^2}{2R_{gg}R^2}e^{R_{gg}z_h^2}).
\end{equation}
The original paper~\cite{Bohra:2019ebj} assumes that the dilaton field $\phi_0$ remains real everywhere in the bulk, which leads to magnetic field $B\leq B_c\simeq0.61GeV$. In this holographic inverse magnetic catalysis model, the deconfinement temperature is $T_c=0.268GeV$ at zero chemical potential and magnetic field.

\section{Spectral function}\label{sec:03}
In this section, we will calculate the spectral function for the $J/\psi$ state and $\Upsilon(1S)$ state by a phenomenological model proposed in Ref.~\cite{Braga:2017bml}. The vector field $A_m=(A_\mu, A_z)(\mu=0,1,2,3)$ is used to represent the heavy quarkonium, which is dual to the gauge theory current $J^\mu=\overline{\Psi}\gamma^\mu\Psi$. The standard Maxwell action takes the following form
\begin{equation}\label{eq06}
  S=\int d^4xdz\sqrt{-g}e^{-\phi(z)}[-\frac{1}{4{g_5}^2}F_{mn}F^{mn}],
\end{equation}
where $F_{mn}=\partial_mA_n-\partial_nA_m$. The scalar field $\phi(z)$, including three energy parameters, is used to parameterize different mesons,
\begin{equation}\label{eq07}
  \phi(z)=\kappa^2z^2+Mz+\tanh(\frac{1}{Mz}-\frac{\kappa}{\sqrt{\Gamma}}),
\end{equation}
where $\kappa$ labels the quark mass, $\Gamma$ is the string tension of the quark pair and $M$ denotes a large mass related to the heavy quarkonium non-hadronic decay. There is a matrix element $\langle0|J_\mu(0)|X(1S)\rangle=\epsilon_\mu f_n m_n$ ($X$ shows the heavy mesons $\Upsilon$ or $\Psi$, $\langle0|$ is the hadronic vacuum, $J_\mu$ is the hadronic current and $f_n$ indicates the decay constant)when the heavy vector mesons decay into leptons. The value of three energy parameters for charmonium and bottomonium in the scalar field, determined by fitting the spectrum of masses~\cite{Braga:2018zlu}, are respectively:
\begin{align}\label{eq08}
  \kappa_c & =1.2GeV,\quad \sqrt{\Gamma_c}=0.55GeV,\quad M_c=2.2GeV,\notag \\
  \kappa_b & =2.45GeV,\quad \sqrt{\Gamma_b}=1.55GeV,\quad M_b=6.2GeV.
\end{align}

The spectral function for $J/\Psi$ state and $\Upsilon(1S)$ state will be calculated with the help of the membrane paradigm~\cite{Iqbal:2008by}. The equation of motion obtained from Eq.~\eqref{eq06} are as follows
\begin{equation}\label{eq09}
  \partial_m(\sqrt{-g}e^{-\phi(z)}F^{mn})=0.
\end{equation}
For the z-foliation, the conjugate momentum of the gauge field $A^\mu$ is given by the following formula:
\begin{equation}\label{eq10}
 j^\mu=-\sqrt{-g}e^{-\phi(z)}F^{z\mu}.
\end{equation}
Supposing plane wave solutions for $A^\mu$ have nothing to do with the coordinates $x_2$ and $x_3$-in other words, the plane wave propagates in the $x_1$ direction. The equation of motion~\eqref{eq09} can be split into two parts: longitudinal-the fluctuations along $(t,x_1)$; transverse-fluctuations along ($x_2,x_3$). Combined with Eq.\eqref{eq10}, the longitudinal components $t, x_1, z$ of Eq.\eqref{eq09} can be expressed respectively as
\begin{align}
  -\partial_zj^t-\frac{R\sqrt{S(z)}e^{B^2z^2-\phi}}{zf(z)}\partial_{x_1}F_{x_1t} & =0, \label{eq11}\\
  -\partial_zj^{x_1}+\frac{R\sqrt{S(z)}e^{B^2z^2-\phi}}{zf(z)}\partial_tF_{x_1t} & =0,\label{eq12} \\
  \partial_{x_1}j^{x_1}+\partial_tj^t & =0\label{eq13}.
\end{align}
\begin{table}
	\centering  
	\begin{tabular}{II c II c I c I c I c I}
		\toprule[1pt]
		& & & &  \\[-6pt]  
		$n$  &$\textbf{State}$    &$\textbf{M}_\textbf{{exp}}\textbf{(MeV)}$    &\textbf{model1(MeV) }  &\textbf{model2(MeV)} \\  
        \midrule[1pt]
		\midrule[1pt]
		& & & &  \\[-6pt]  
		1&$J/\Psi$&$3096.916\pm0.011$&3037.1&3510.7 \\
		
		& & & &  \\[-6pt]  
		2&$\Psi(2S)$&$3686.109\pm0.012$&4126.0&4082.2 \\

        & & & &  \\[-6pt]  
		3&$\Psi(3S)$&$4039\pm1$&4971.9 &4523.2 \\

        & & & &  \\[-6pt]  
		4&$\Psi(4S)$&$4421\pm4$&5679.0 &4890.8 \\
        \bottomrule[1pt]
	\end{tabular}
\caption{Holographic masses for the Charmonium S-wave resonances. Experimental results are from PDG\cite{ParticleDataGroup:2018ovx}.}  
	\label{tab1}  
\end{table}
\begin{table}
	\centering  
	\begin{tabular}{II c II c I c I c I c I}
		\toprule[1pt]
		& & & &  \\[-6pt]  
		$n$  &$\textbf{State}$    &$\textbf{M}_\textbf{{exp}}\textbf{(MeV)}$    &\textbf{model1(MeV) }  &\textbf{model2(MeV)} \\  
        \midrule[1pt]
		\midrule[1pt]
		& & & &  \\[-6pt]  
		1&$\Upsilon(1S)$&$9460.3\pm0.26$&7073.7&9196.0 \\
		
		& & & &  \\[-6pt]  
		2&$\Upsilon(2S)$&$10023.26\pm0.32$&9040.4&10276.6 \\

        & & & &  \\[-6pt]  
		3&$\Upsilon(3S)$&$10355.2\pm0.5$&10620.0&10645.9 \\

        & & & &  \\[-6pt]  
		4&$\Upsilon(4S)$&$10579.4\pm1.2$&11963.0 &10951.9 \\
        \bottomrule[1pt]
	\end{tabular}
\caption{Holographic masses for the Bottomonium S-wave resonances. Experimental results are from PDG\cite{ParticleDataGroup:2018ovx}.}  
	\label{tab2}  
\end{table}
By using the Bianchi identity, one can get
\begin{equation}\label{eq14}
  \partial_zF_{x_1t}-\frac{ze^{\phi-B^2z^2}}{R\sqrt{S(z)}f(z)}\partial_tj^{x_1}-\frac{zf(z)e^{\phi-B^2z^2}}{R\sqrt{S(z)}}\partial_{x_1}j^t=0.
\end{equation}
The longitudinal "conductivity" and its derivative are defined as
\begin{align}
  \sigma_L(\omega,\overrightarrow{p},z) & =\frac{j^{x_1}(\omega,\overrightarrow{p},z)}{F_{x_1t}(\omega,\overrightarrow{p},z)},\label{eq15} \\
  \partial_z \sigma_L(\omega,\overrightarrow{p},z) & =\frac{\partial_zj^{x_1}}{F_{x_1t}}-\frac{j^{x_1}}{F_{x_1t}^2}\partial_zF_{x_1t}\label{eq16}.
\end{align}
The transverse channel is decided by following the dynamical equation
\begin{align}
  -\partial_zj^{x_2}-\frac{Re^{-\phi}\sqrt{S(z)}}{z}(\frac{\partial_tF_{tx_2}}{f(z)}-\partial_{x_1}F_{x_1x_2}) & = 0,\label{eeq1}\\
  \partial_zF_{x_2t}-\frac{ze^\phi}{R\sqrt{S(z)}f(z)}\partial_tj^{x_2} & = 0,\label{eeq2}\\
  \partial_{x_1}F_{tx_2}+\partial_tF_{x_2x_1} & = 0.\label{eeq3}
\end{align}
The latter two equations from the Bianchi identity are constraint equations. Now, the transverse "conductivity" and its derivative are defined as
\begin{align}
  \sigma_T(\omega,\overrightarrow{p},z) & =\frac{j^{x_2}(\omega,\overrightarrow{p},z)}{F_{x_2t}(\omega,\overrightarrow{p},z)},\label{eeq4} \\
  \partial_z \sigma_T(\omega,\overrightarrow{p},z) & =\frac{\partial_zj^{x_2}}{F_{x_2t}}-\frac{j^{x_2}}{F_{x_2t}^2}\partial_zF_{x_2t}\label{eeq5}.
\end{align}
The Kubo's formula shows that the five-dimensional "conductivity" at the boundary is related to the retarded Green's function:
\begin{align}
  \sigma_L(\omega) & =\frac{-G_R^L(\omega)}{i\omega},\label{eeq6}\\
  \sigma_T(\omega) & =\frac{-G_R^T(\omega)}{i\omega},\label{eeq7}
\end{align}
 where $\sigma_L$ is interpreted as the longitudinal AC conductivity and  $\sigma_T$ denotes the transverse AC conductivity; Here $G_R^L(\omega)$ and $G_R^T(\omega)$ are longitudinal and transverse retarded two-point Green's function of the currents, which can be calculated from the solutions of equations of motion from AdS/CFT~\cite{Iqbal:2008by,Son:2002sd}. To obtain flow equations ~\eqref{eq16} and ~\eqref{eeq5}, we assume $A_\mu=A_n(p,z)e^{-i\omega t+ipx_1}$, where $A_n(p,z)$ is the quasinormal modes. Therefore, we have $\partial_tF_{x_1t}=-i\omega F_{x_1t}$, $\partial_tj^{x_1}=-i\omega j^{x_1}$ for Eq. ~\eqref{eq16} and $\partial_{t}F_{x_1x_2}=-i\omega F_{x_1x_2}$,\,$\partial_tF_{tx_2}=-i\omega F_{tx_2}$,\,$\partial_tj^{x_2}=-i\omega j^{x_2}$ for Eq.\eqref{eeq5} .
 By combining Eq.\eqref{eq12}, Eq.\eqref{eq13}, Eq.\eqref{eq14} for longitudinal case and Eq.\eqref{eeq1}, Eq.\eqref{eeq2}, Eq.\eqref{eeq3} for the transverse case and the momentum $P=(\omega,0,0,0)$ with metric\eqref{eq02}, the Eq.\eqref{eq16} and Eq.\eqref{eeq5} can be written as
\begin{align}
 \partial_z \sigma_L & =\frac{iz\omega}{f(z)e^{B^2z^2-\phi+A(z)}}(\sigma_L^2-\frac{e^{2(B^2z^2-\phi+A(z))}}{z^2}),\label{eq18}\\
 \partial_z \sigma_T & =\frac{iz\omega}{f(z)e^{A(z)-\phi}}(\sigma_T^2-\frac{e^{2(A(z)-\phi)}}{z^2}).\label{eq19}
\end{align}
The initial condition for solving the equation can be obtained by  requiring regularity at the horizon $\partial_z\sigma_{L(T)}=0$. It is not difficult to find that the metric Eq.\eqref{eq02} restores the SO(3) invariance and the flow equations Eq.\eqref{eq18} and Eq.\eqref{eq19} have the same form when magnetic field $B=0GeV$.
The spectral function is defined by the retarded Green's function
\begin{equation}\label{eq20}
  \rho(\omega)\equiv-Im G_R(\omega)=\omega Re\,\sigma|_{z=0}.
\end{equation}
\section{Conclusion and discussion}\label{sec:04}
In this paper, By making use of  a general magnetic field dependent dynamical AdS/QCD model
we  investigate the spectral function of heavy mesons($J/\Psi$ and $\Upsilon(1S)$). Here, the heavy vector mesons are represented by a phenomenological model proposed in Ref.~\cite{Braga:2017bml}. The spectral function are calculated by using the membrane paradigm~\cite{Iqbal:2008by}. First, we show the holographic quarkonium masses in zero temperature case (see \cite{Karch:2006pv} for the detailed calculation process), in tables \ref{tab1} and \ref{tab2}, for the states $1S, 2S, 3S, 4S$ in two different dilatons, quadratic dilaton(model1)-$\phi(z)=\kappa^2z^2+Mz+\tanh(\frac{1}{Mz}-\frac{\kappa}{\sqrt{\Gamma}})$, nonquadratic dilaton(model2)~\cite{MartinContreras:2021bis}-$\phi(z)=(\kappa z)^{2-\alpha}+Mz+\tanh(\frac{1}{Mz}-\frac{\kappa}{\sqrt{\Gamma}})$, for the values of  parameters in model2, see Ref.~\cite{MartinContreras:2021bis}. Further, we compare the theoretical values with the experimental values\cite{ParticleDataGroup:2018ovx} for heavy mesons mass.
\begin{figure}
  \centering
  \begin{minipage}[t]{0.23\textwidth}
  \centering
  \includegraphics[width=4.2cm]{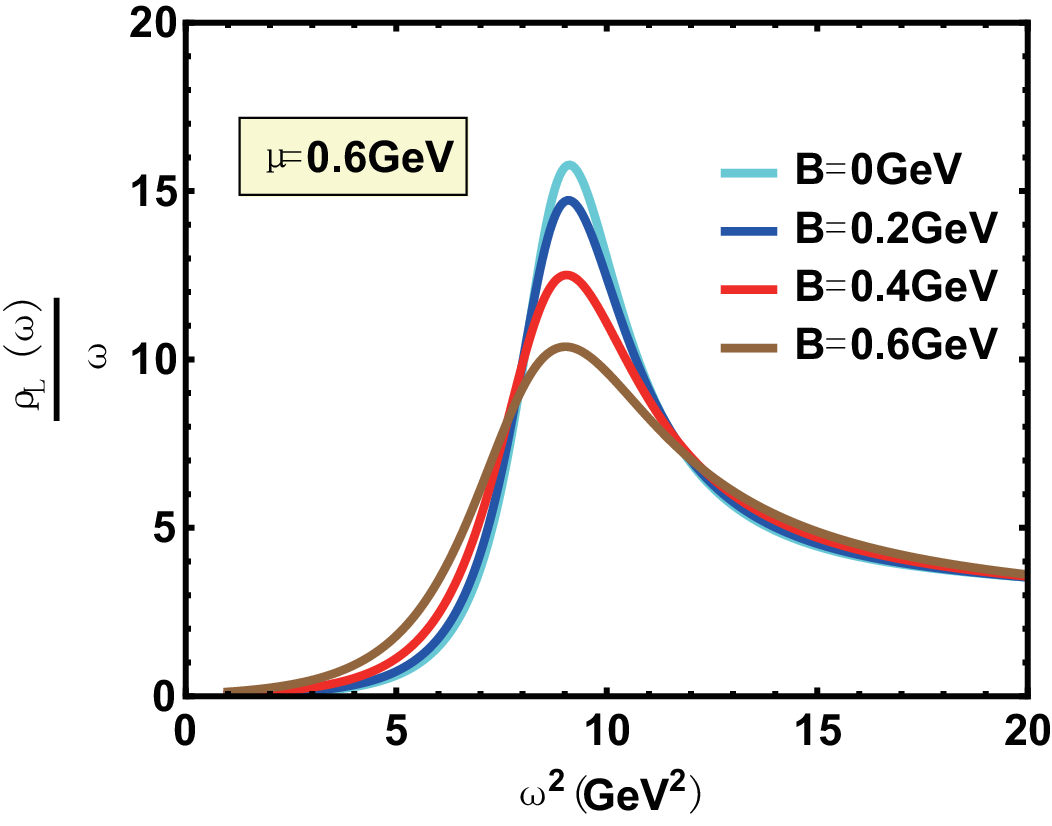}\\
  \end{minipage}
  \begin{minipage}[t]{0.23\textwidth}
  \centering
  \includegraphics[width=4.2cm]{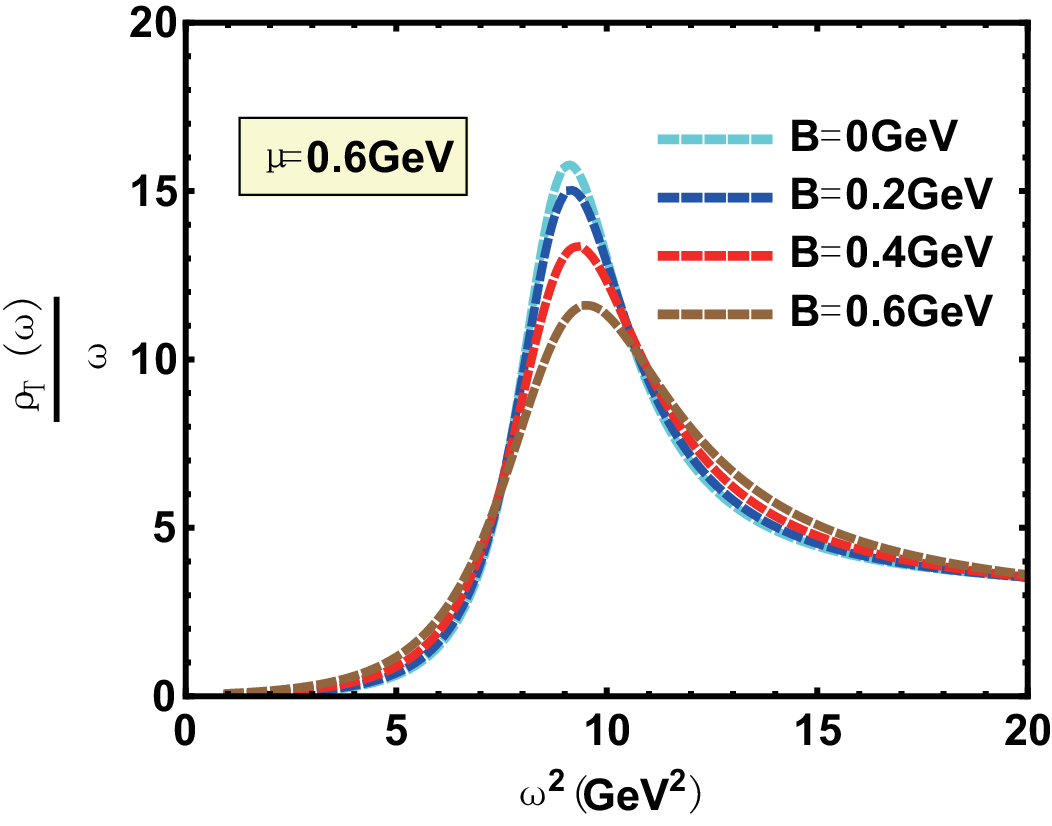}\\
  \end{minipage}
  \caption{The spectral function of the $J/\Psi$ state with different magnetic field $B$ at $T=0.3GeV$. The left picture is for magnetic field parallel to polarization and the right one is for magnetic field perpendicular to polarization. }\label{fig1}
\end{figure}
\begin{figure}
  \centering
  \begin{minipage}[t]{0.23\textwidth}
  \centering
  \includegraphics[width=4.2cm]{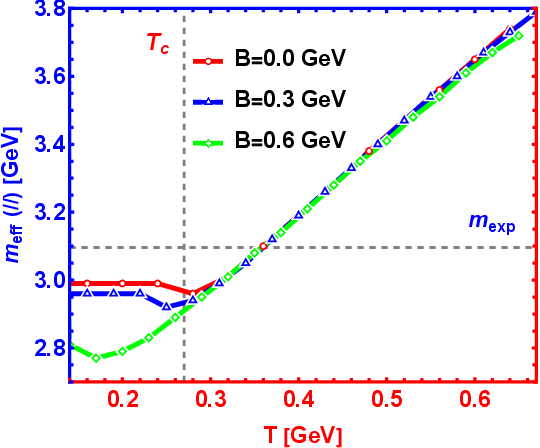}\\
  \end{minipage}
  \begin{minipage}[t]{0.23\textwidth}
  \centering
  \includegraphics[width=4.2cm]{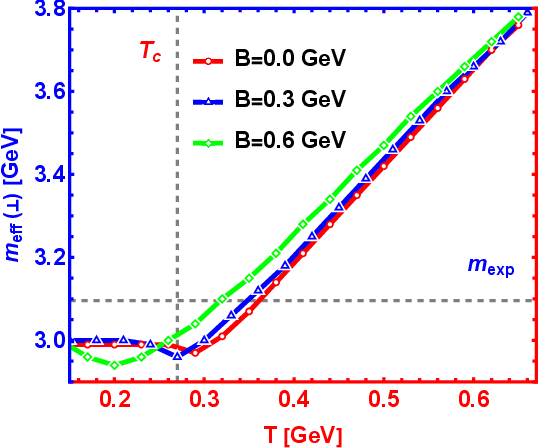}\\
  \end{minipage}
  \caption{The effective mass of $J/\Psi$ state  for different magnetic field $B$ at $\mu=0GeV$ . The left picture is for magnetic field parallel to polarization and the right one is for magnetic field perpendicular to polarization.}\label{mBpsi}
\end{figure}
\begin{figure}
  \centering
  \begin{minipage}[t]{0.23\textwidth}
  \centering
  \includegraphics[width=4.2cm]{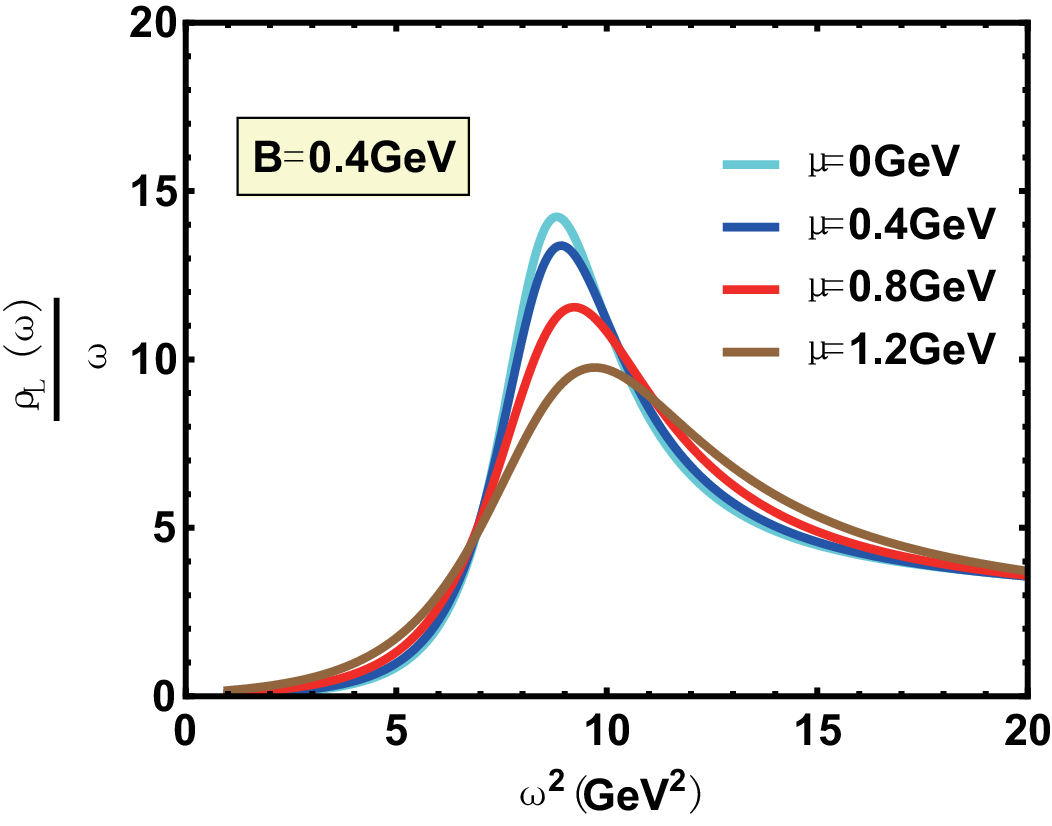}\\
  \end{minipage}
  \begin{minipage}[t]{0.23\textwidth}
  \centering
  \includegraphics[width=4.2cm]{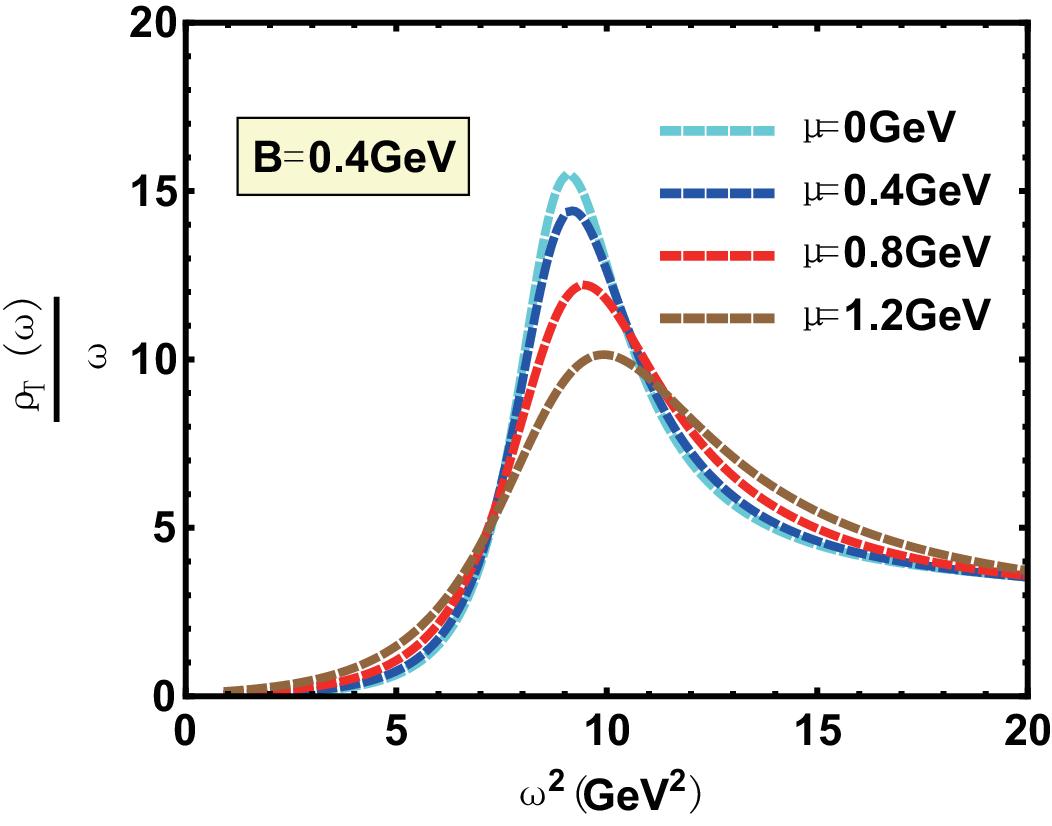}\\
  \end{minipage}
  \caption{The spectral function of the $J/\Psi$ state with different chemical potential $\mu$ at $T=0.3GeV$. The left picture is for magnetic field parallel to polarization and the right one is for magnetic field perpendicular to polarization. }\label{fig2}
\end{figure}
\begin{figure}
  \centering
  \begin{minipage}[t]{0.23\textwidth}
  \centering
  \includegraphics[width=4.2cm]{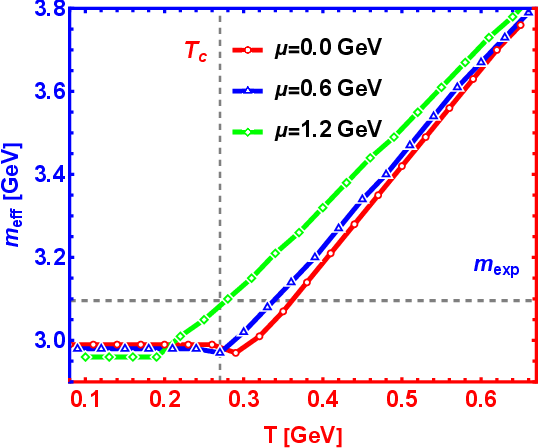}\\
  \end{minipage}
  \begin{minipage}[t]{0.23\textwidth}
  \centering
  \includegraphics[width=4.2cm]{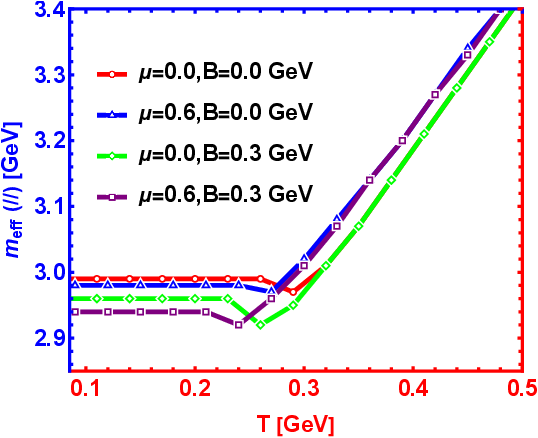}\\
  \end{minipage}
  \caption{Left: The effective mass of $J/\Psi$ state  for different chemical potential $\mu$ at $B=0 GeV$. Right: The effective mass of $J/\Psi$ state for the simultaneous effects of $B$ and $\mu$ in the case of magnetic field parallel to polarization.}\label{mmupsi}
\end{figure}
\begin{figure}
  \centering
  \begin{minipage}[t]{0.23\textwidth}
  \centering
  \includegraphics[width=4.2cm]{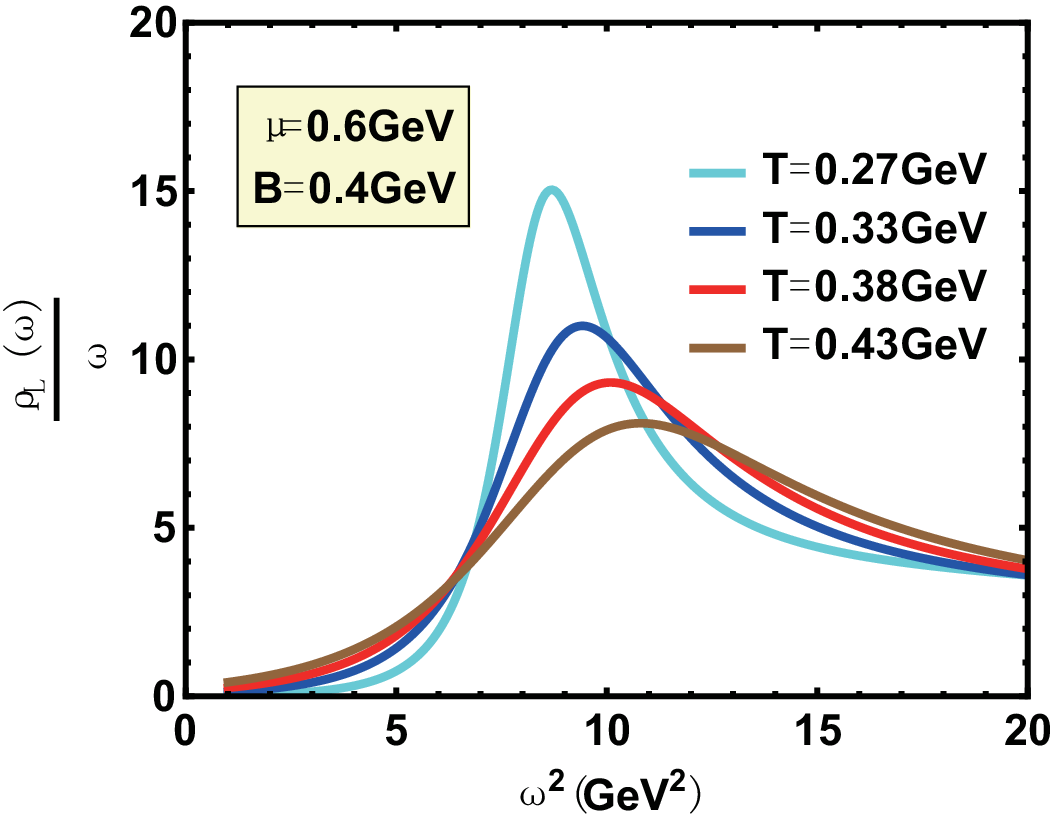}\\
  \end{minipage}
  \begin{minipage}[t]{0.23\textwidth}
  \centering
  \includegraphics[width=4.2cm]{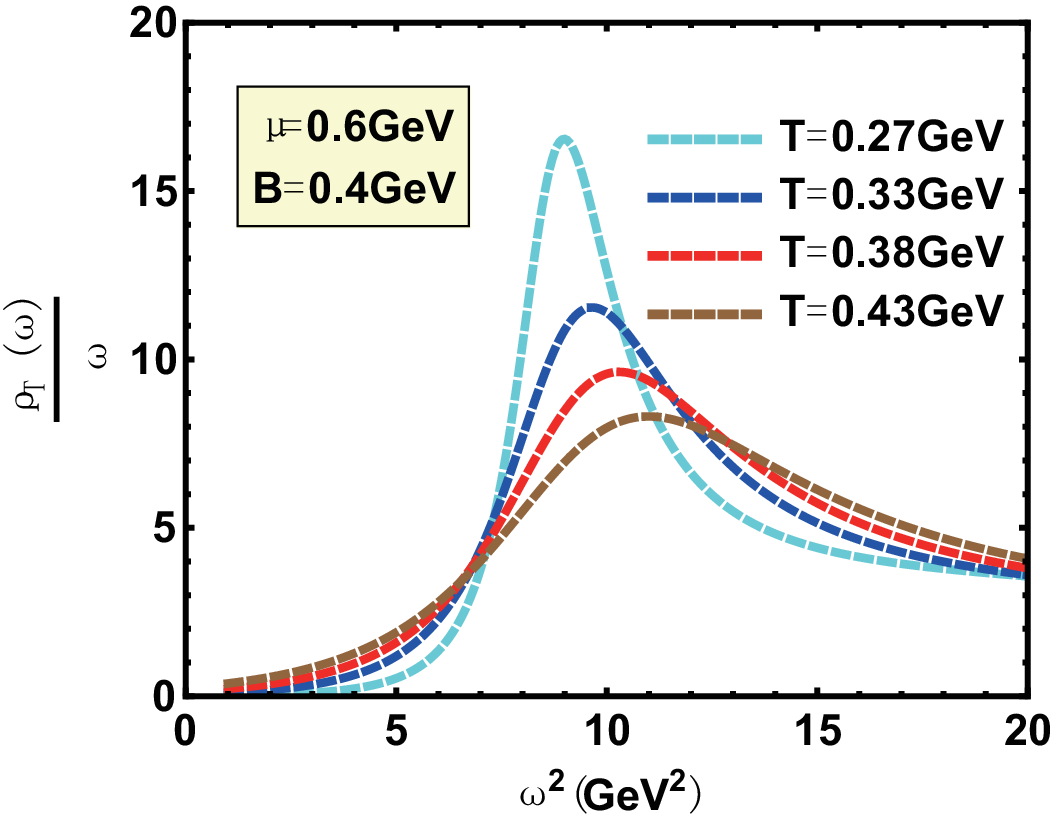}\\
  \end{minipage}
  \caption{The spectral function of the $J/\Psi$ state  for different temperature $T$ . The left picture is for magnetic field parallel to polarization and the right one is for magnetic field perpendicular to polarization. }\label{fig3}
\end{figure}
\begin{figure}
  \centering
  \begin{minipage}[t]{0.23\textwidth}
  \centering
  \includegraphics[width=4.3cm]{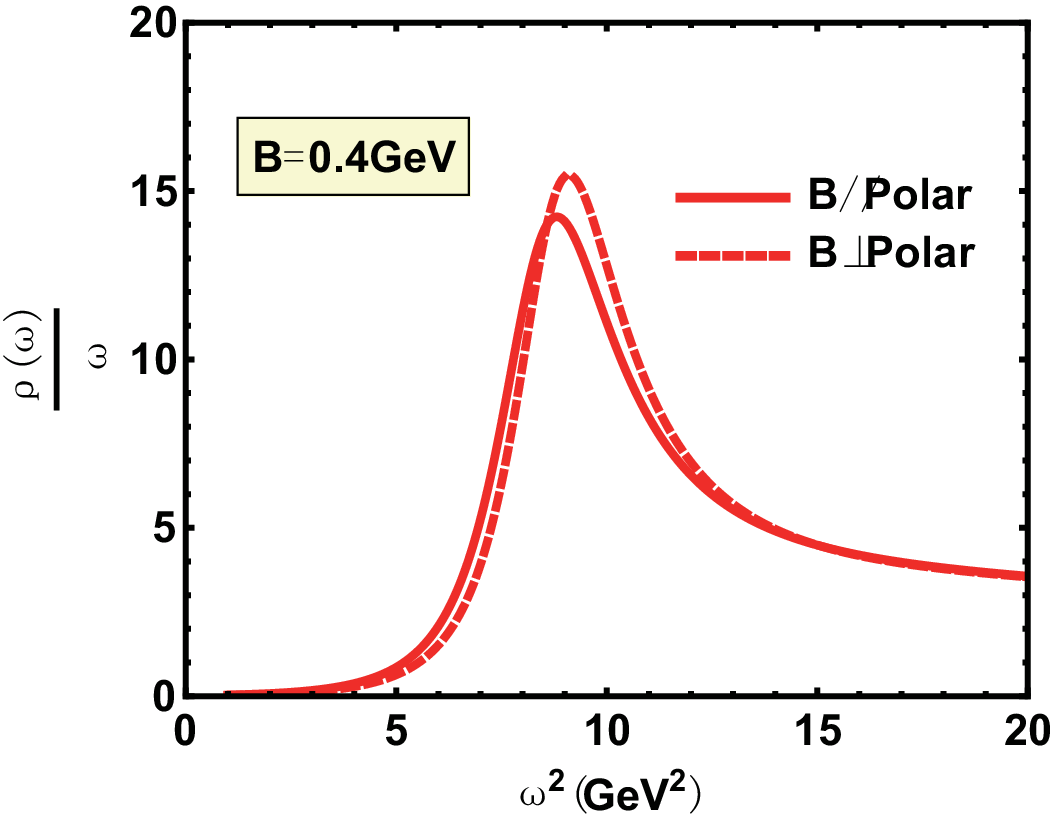}\\
  \end{minipage}
  \begin{minipage}[t]{0.23\textwidth}
  \centering
  \includegraphics[width=3.95cm]{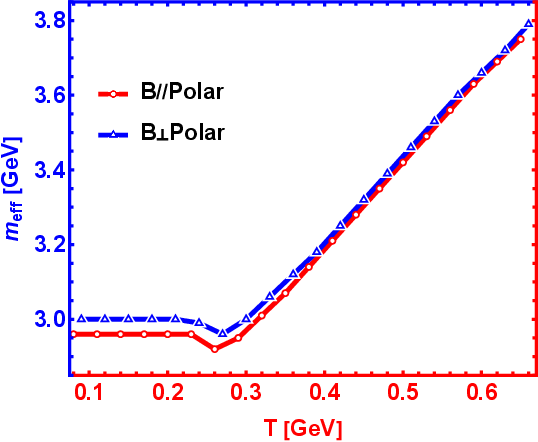}\\
  \end{minipage}
  \caption{Left: The spectral function of the $J/\Psi$ state for parallel case and perpendicular case at $T=0.3GeV, \mu=0GeV$. Right: The effective mass of the $J/\Psi$ state for parallel case and perpendicular case at $B=0.3GeV, \mu=0GeV$.}\label{fig13}
\end{figure}

\subsection{The Result For $J/\Psi$}
In Fig.\ref{fig1}, the spectral function of the $J/\Psi$ state for different magnetic fields parallel and perpendicular to the polarization are shown. The bell shape represents the resonance state and peak position $\omega$ is related to the resonant mass. The width corresponding to half the peak height is the resonant decay width. One can find that increasing magnetic field reduces the height of the peak and enlarges its width of that. This phenomenon indicates that the appearance of magnetic field speeds the dissociation which is consistent with Ref.\cite{Braga:2018zlu} where the authors use an Einstein-Maxwell theory. In addition, an interesting phenomenon about effective mass is shown in Fig.\ref{mBpsi}. When magnetic field is parallel to the polarization, the magnetic field reduces the effective mass, reminiscent of the inverse magnetic catalysis at lower temperature, while that is from inverse magnetic catalysis to magnetic catalysis in ref.\cite{Braga:2019yeh} where the authors use an Einstein-Maxwell theory.  It is worth noting that, after phase transition temperature, our results show the effective mass is almost independent of magnetic field with only slight suppresion, however, the magnetic field presents a magnetic catalysis effect in ref.\cite{Braga:2019yeh}. When magnetic field is perpendicular to the polarization, our conclusions give that the magnetic field effect on the effective mass is from magnetic catalysis to inverse magnetic catalysis at lower temperature and behaves as magnetic catalysis after phase transition temperature, however, an inverse magnetic catalysis effect is obtained for all temperature in Ref.\cite{Braga:2019yeh}. Here, we need to point out that in this paper, the effective mass is determined by the energy $\omega$ corresponding to the position of the spectral function peak, while in Ref.\cite{Braga:2019yeh}, it is determined by the real part of the quasinormal modes.

We present the spectral function of the $J/\Psi$ state for different chemical potentials in Fig.\ref{fig2}. It can be found that chemical potential has a similar effect as the magnetic field. Increasing chemical potential reduces the height of the peak and increases the width which is consistent with  Ref.\cite{Braga:2017bml} where the authors use a generalized soft-wall model. However, attention must be paid to the influence of chemical potential on the peak position, which is plotted in the left picture of Fig.\ref{mmupsi}. The result suggests that the chemical potential gives a suppression effect at lower temperature and presents a enhancement effect at higher temperature on the effective mass, the same conclusion is obtained in Ref.\cite{Braga:2019xwl} where the authors use a generalized soft-wall model. However we find that chemical potential effect is more prominent at higher temperature in this paper, while  the results from Ref.\cite{Braga:2019xwl} shows that the chemical potential effect  mainly acts at lower temperature. Besides, the simultaneous effects of $B$ and $\mu$ is depicted in the right picture of Fig.\ref{mmupsi}. The superposition of chemical potential and magnetic field effects causes non-trivial  behaviors of the effective mass :  At lower temperature, there appears constructive interplay on the effective mass between the effects of chemical potential and magnetic field , while at  higher temperature, the chemical potential and magnetic field  lead to  destructive interplay on the effective mass.

In Fig.\ref{fig3}, the behavior of spectral function is extraordinarily sensitive to the temperature
of the current physical system. With the increase of the medium temperature, the height of the peak decreases rapidly which is in line with Ref.\cite{Braga:2016wkm}. The effect of temperature on the width and the position of the peak (see Fig.\ref{mBpsi} and \ref{mmupsi}) is similar to that of chemical potential, which is  consistent with Ref.\cite{MartinContreras:2021bis} and  the lattice results in Ref.\cite{Ding:2018uhl}.

Besides, in Fig.\ref{fig13}, we compare the magnetic field parallel to polarization with perpendicular to that. The comparison shows that the melting effect produced by the magnetic field is stronger and the effective mass is smaller in the parallel direction than that in the perpendicular direction, which is diametrically opposite to that found in Ref.\cite{Braga:2018zlu} in the EM model. Our result agrees qualitatively with the lattice result of string tension in Ref.~\cite{Bonati:2016kxj}, where it is found that the transverse magnetic field increases the string tension , while the longitudinal magnetic field suppresses the string tension. This difference comes from two aspects: One is this result can be well explained by comparing Eq.\eqref{eq18} and Eq.\eqref{eq19} in terms of appearance, the appearance of the magnetic field in the denominator of the first term on the right of Eq.~\eqref{eq18} and  the numerator inside the bracket both decrease the value of $Re\,\sigma$ numerically. In contrast, Eq.\eqref{eq19} does not have a magnetic field, formally. The second is the dilaton introduced in the action, in other words, that is the difference between the EM model and the EMD model.  In this paper, we consider the EMD model. The advantages of EMD model can be found in Ref.~\cite{Zhou:2020ssi}. The dilaton will deform the space and influence the magnetic field in different directions. Further, we find EMD model can realize the linear part of Cornell potential extracted from lattice simulation and show the magnetic field in a parallel direction has a large influence on potential. The EM action without the dilaton field cannot get a proper free energy behavior of single quark. The result from lattice ~\cite{Bonati:2016kxj} also favors the EMD model with a dilaton than the EM model.
\subsection{The Result For $\Upsilon(1S)$}
\begin{figure}[htbp]
  \centering
  \begin{minipage}[t]{0.23\textwidth}
  \centering
  \includegraphics[width=4.2cm]{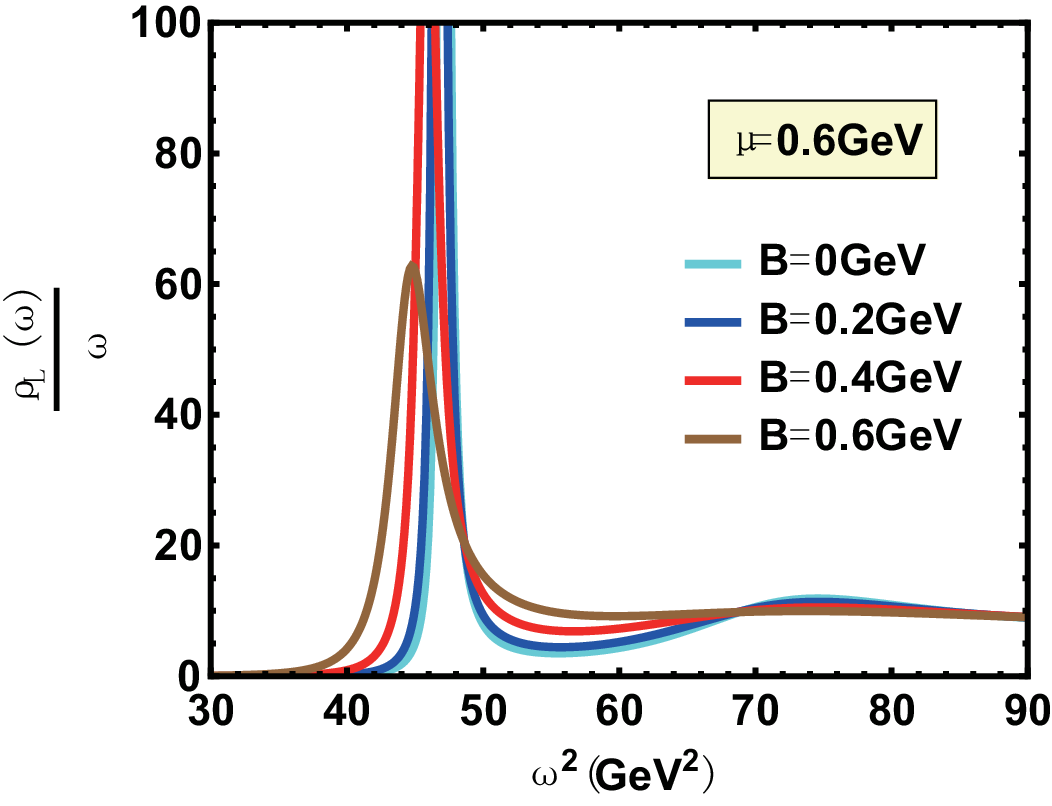}\\
  \end{minipage}
  \begin{minipage}[t]{0.23\textwidth}
  \centering
  \includegraphics[width=4.2cm]{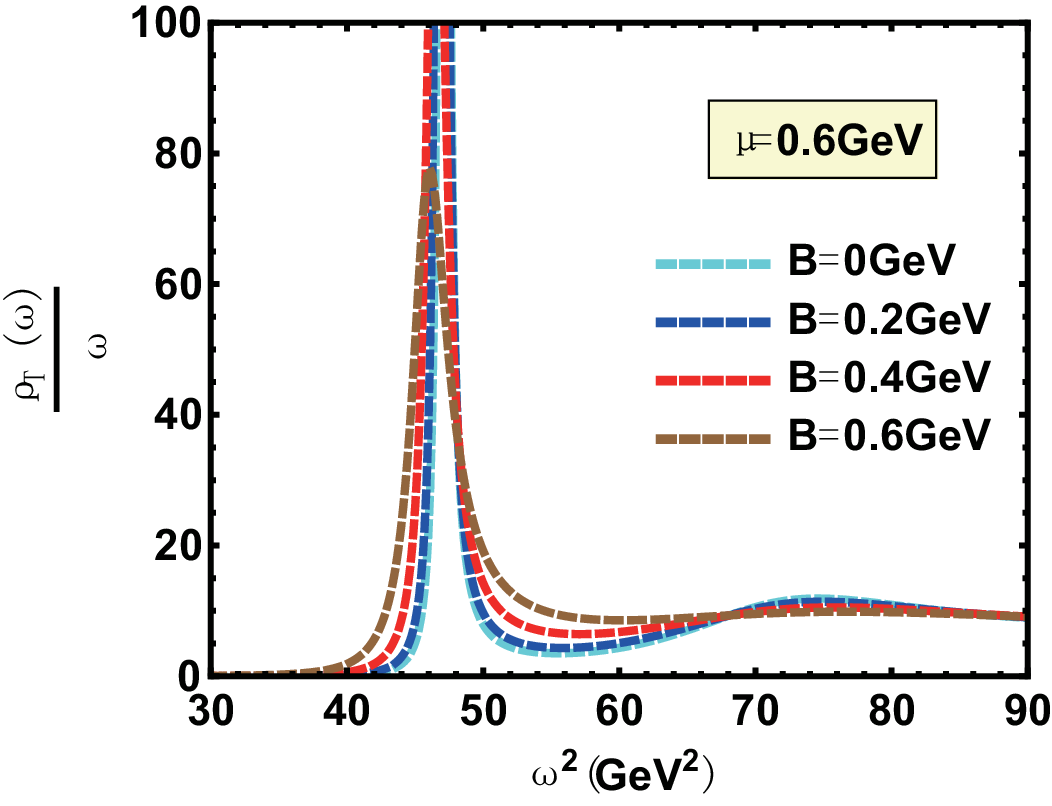}\\
  \end{minipage}
  \caption{The spectral function of the $\Upsilon(1S)$ state with different magnetic field $B$ at $T=0.3GeV$. The left picture is for magnetic field parallel to polarization and the right one is for magnetic field perpendicular to polarization. }\label{fig7}
\end{figure}
\begin{figure}
  \centering
  \begin{minipage}[t]{0.23\textwidth}
  \centering
  \includegraphics[width=4.2cm]{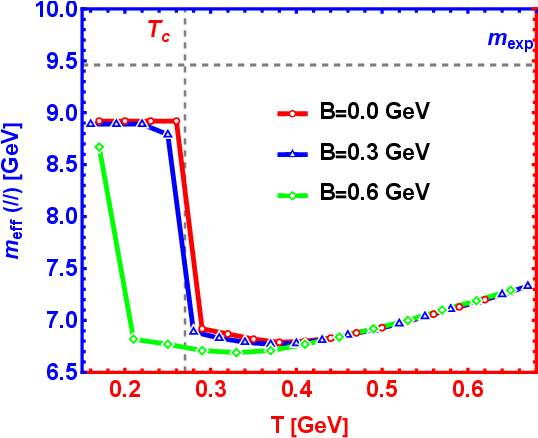}\\
  \end{minipage}
  \begin{minipage}[t]{0.23\textwidth}
  \centering
  \includegraphics[width=4.2cm]{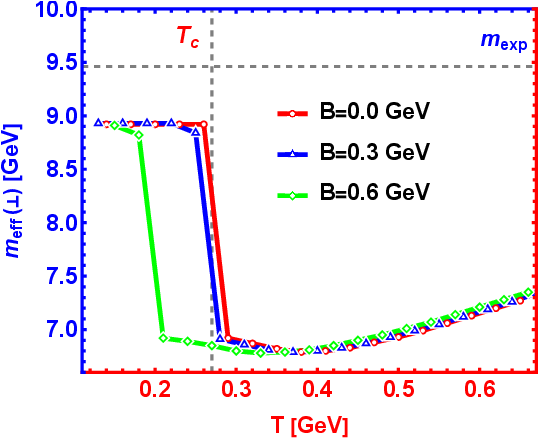}\\
  \end{minipage}
  \caption{The effective mass of $\Upsilon(1S)$ state  for different magnetic field $B$ at $\mu=0GeV$ . The left picture is for magnetic field parallel to polarization and the right one is for magnetic field perpendicular to polarization.}\label{mBupsilon}
\end{figure}
\begin{figure}[htbp]
  \centering
  \begin{minipage}[t]{0.23\textwidth}
  \centering
  \includegraphics[width=4.2cm]{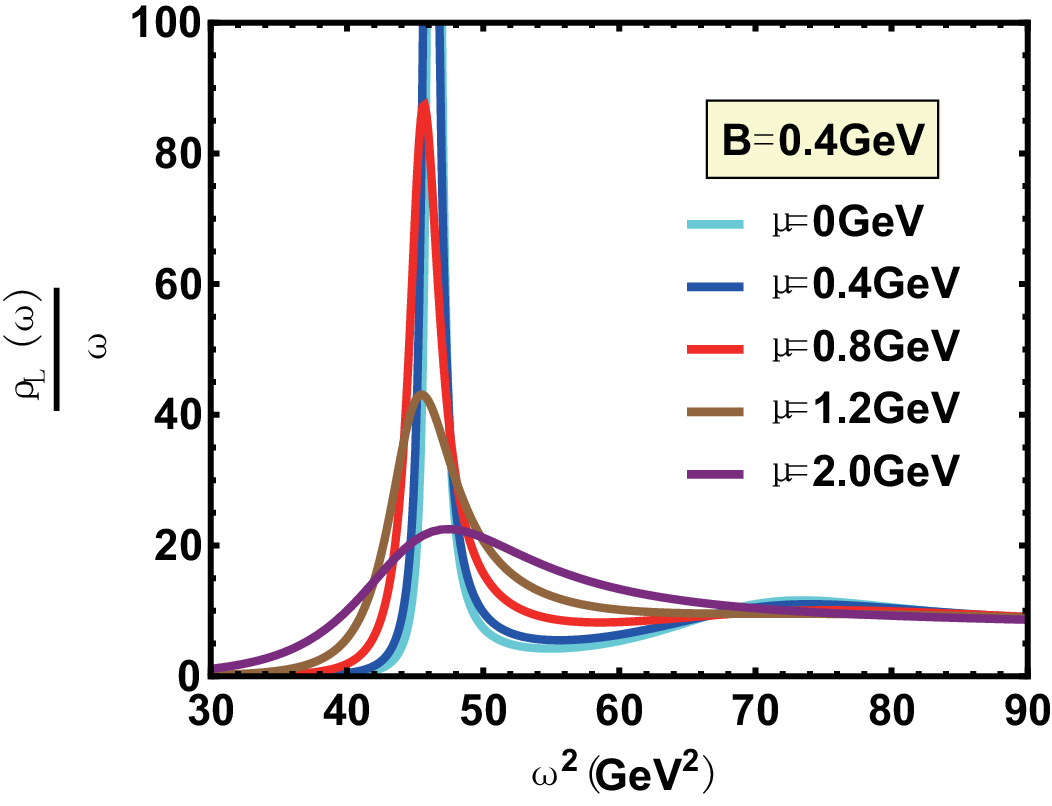}\\
  \end{minipage}
  \begin{minipage}[t]{0.23\textwidth}
  \centering
  \includegraphics[width=4.2cm]{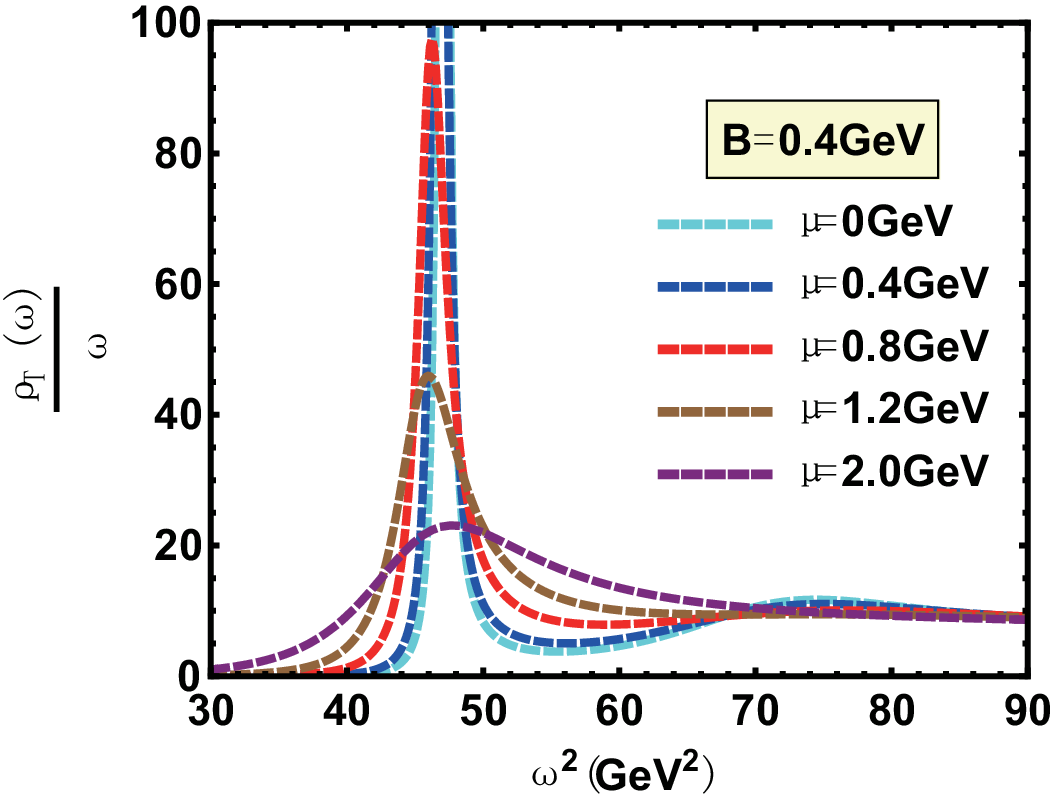}\\
  \end{minipage}
  \caption{The spectral function of the $\Upsilon(1S)$ state with different chemical potential $\mu$ at $T=0.3GeV$. The left picture is for magnetic field parallel to polarization and the right one is for magnetic field perpendicular to polarization. }\label{fig8}
\end{figure}
\begin{figure}
  \centering
  \begin{minipage}[t]{0.23\textwidth}
  \centering
  \includegraphics[width=4.2cm]{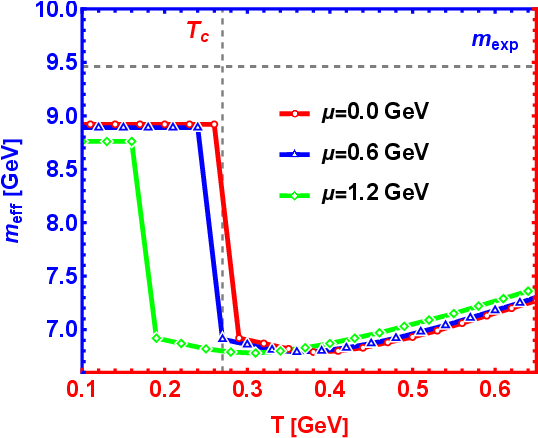}\\
  \end{minipage}
  \begin{minipage}[t]{0.23\textwidth}
  \centering
  \includegraphics[width=4.2cm]{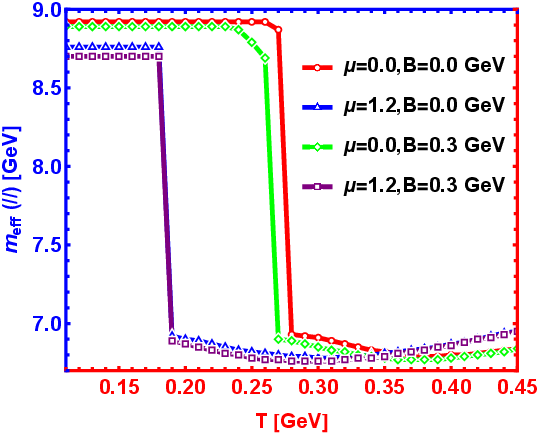}\\
  \end{minipage}
  \caption{Left: The effective mass of $\Upsilon(1S)$ state  for different chemical potential $\mu$ at $B=0 GeV$. Right: The effective mass of $\Upsilon(1S)$ state for the simultaneous effects of $B$ and $\mu$ in the case of magnetic field parallel to polarization. }\label{mmuupsilon}
\end{figure}
\begin{figure}[htbp]
  \centering
  \begin{minipage}[t]{0.23\textwidth}
  \centering
  \includegraphics[width=4.2cm]{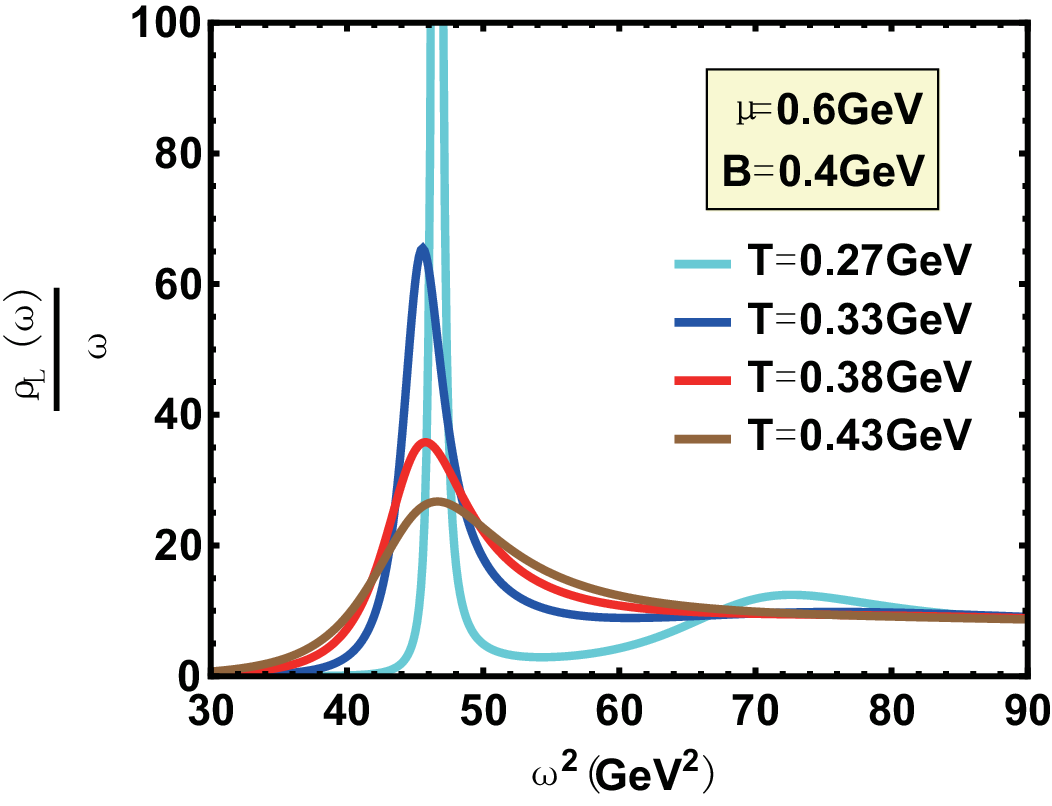}\\
  \end{minipage}
  \begin{minipage}[t]{0.23\textwidth}
  \centering
  \includegraphics[width=4.2cm]{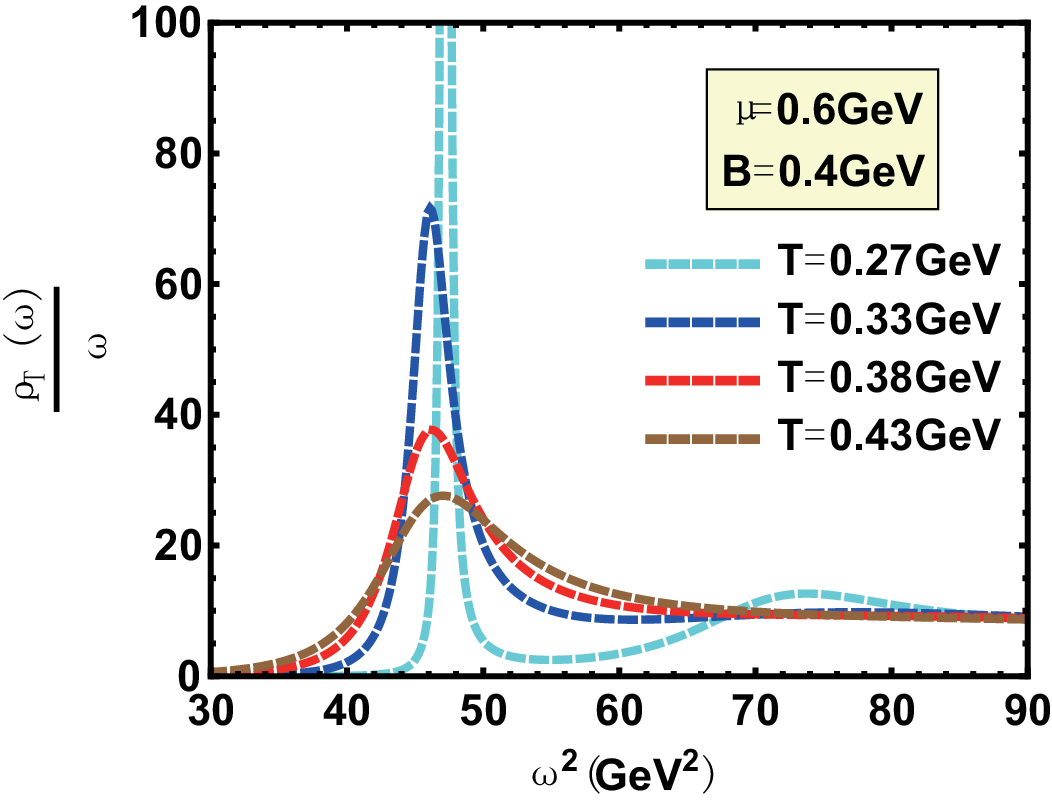}\\
  \end{minipage}
  \caption{The spectral function of the $\Upsilon(1S)$ state  for different temperature $T$ . The left picture is for magnetic field parallel to polarization and the right one is for magnetic field perpendicular to polarization. }\label{fig9}
\end{figure}
\begin{figure}[htbp]
  \centering
  \begin{minipage}[t]{0.23\textwidth}
  \centering
  \includegraphics[width=4.3cm]{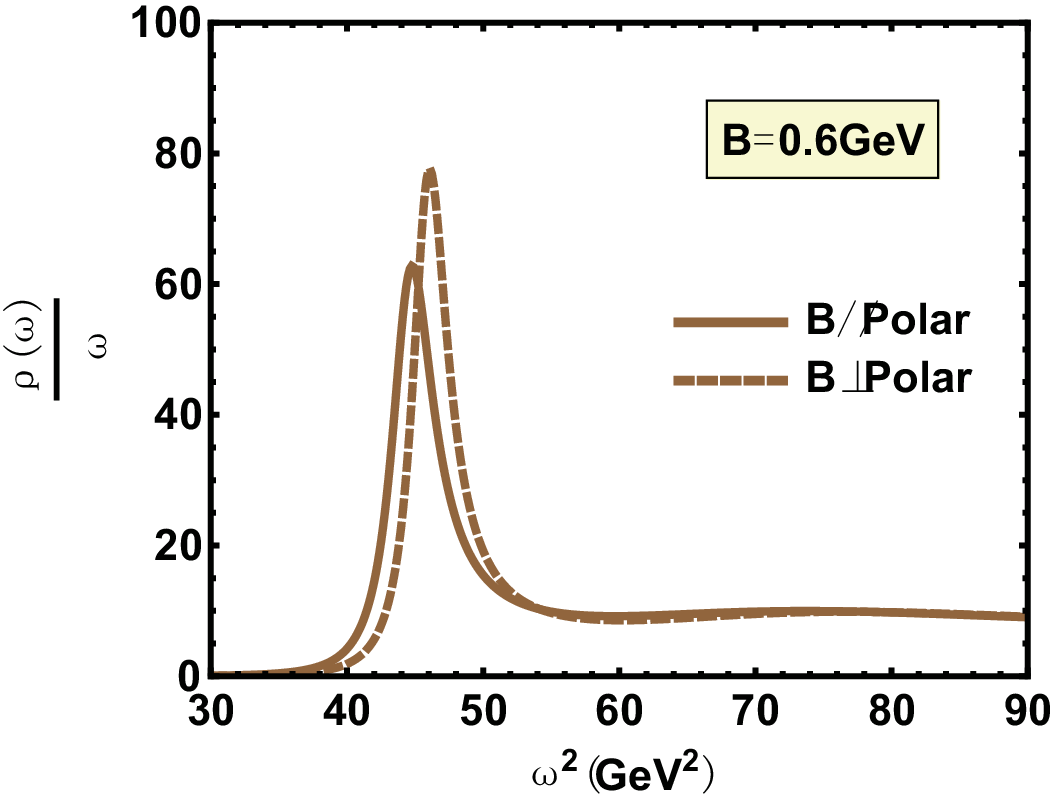}\\
  \end{minipage}
  \begin{minipage}[t]{0.23\textwidth}
  \centering
  \includegraphics[width=3.95cm]{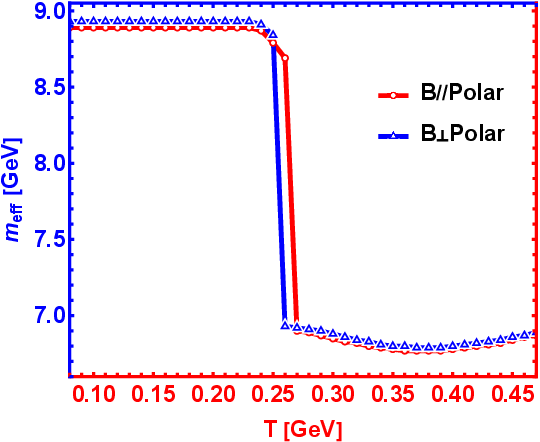}\\
  \end{minipage}
  \caption{Left: The spectral function of the $\Upsilon(1S)$ state for parallel case and perpendicular case at $T=0.3GeV, \mu=0.6GeV$. Right: The effective mass of the $\Upsilon(1S)$ state for parallel case and perpendicular case at $B=0.3GeV, \mu=0GeV$.}\label{fig14}
\end{figure}
The spectral function for the $\Upsilon(1S)$ state with various cases are shown in Fig.~\ref{fig7}-Fig.~\ref{fig14}. Compared with $J/\Psi$ in this paper and $\Upsilon(1S)$ in Ref.\cite{Braga:2019yeh}, one can find the only difference is from the influence of magnetic field on the peak position of spectral function (effective mass). Whether the magnetic field is perpendicular to or parallel to polarization, Fig.\ref{mBupsilon} reports an inverse magnetic catalysis ~\cite{Thakur:2021vbo,Morita:2009qk} at lower temperature and a magnetic catalysis~\cite{Ding:2018uhl,Rothkopf:2016rlu} at higher temperature, which is in line with Ref.\cite{Braga:2019yeh}. The only difference is that the effective mass mutates at $T_c$ for zero magnetic field and the mutation occurs in a smaller temperature with the increasing  magnetic field in our model. The reason for this mutation is that the confinement-deconfinement phase transition here is a first-order phase transition. However, the change of effective mass with temperature is smooth from Ref.\cite{Braga:2019yeh}, which is because the transition from hadron phase to quark phase is crossover.

 For both $J/\Psi$ and  $\Upsilon(1S)$, the superposition of chemical potential and magnetic field effects causes non-trivial  behaviors of the effective mass :  At lower temperature, there appears constructive interplay on the effective mass between the effects of chemical potential and magnetic field , while at  higher temperature, the chemical potential and magnetic field  lead to  destructive interplay on the effective mass. Due to the superposition of two distinct effects between chemical potential and magnetic field, some non-trivial  non monotonic behaviors of effective mass appears in the middle temperature region.  Of course,  the inflection point of this non-trivial behavior strictly depends on the strength of chemical potential, temperature, magnetic field and the interaction between quarkonium. We explain this non-trivial  behavior by the interplay of the interaction between the two heavy quarks and  the interaction between the medium with each of the heavy quarks. Finally, combined with the behavior of spectral function and the influence of superposition effect of $\mu$ and $B$ on effective mass, a general conclusion is obtained, in the strong coupling system, the effective mass is dominated by the depression effect at a lower temperature, smaller chemical potential, weaker magnetic field and shorter distance of interaction; however, the effective mass is dominated by the enhancement effect at a higher temperature, higher chemical potential, stronger magnetic field and longer distance of the heavy quarks. In other words, the effective mass is closely related to the coupling strength and the distance of interaction between the $Q\bar{Q}$ pair.

\section{acknowledgments}
This work is supported by the National Natural Science Foundation of China  by Grant Nos. 11735007,11890711, and 11890710. In addition, the first author wants to thank Hai-cang Ren and  Zhou-Run Zhu for helpful discussions.

\section*{Reference}
\bibliography{ref}

\end{document}